\documentclass[a4paper]{article}
\usepackage[top=1in, bottom=1.25in, left=1.25in, right=1.25in]{geometry}
\usepackage{amsmath,amssymb,amsthm,times}
\usepackage{framed}
\usepackage{tikz}
\usepackage{enumitem}
\usepackage{pdfpages}
\usepackage{mathtools}

\newcommand{\bx}{\mathbf{x}}
\newcommand{\xl}{x_L}
\newcommand{\xr}{x_R}
\newcommand{\xc}{x_C}

\newcommand{\cost}{\mbox{\rm cost}}
\renewcommand{\Re}{\mathbb{R}}

\newtheorem{theorem}{Theorem}
\newtheorem{definition}[theorem]{Definition}
\newtheorem{lemma}[theorem]{Lemma}
\newtheorem{claim}[theorem]{Claim}
\newtheorem{observation}[theorem]{Observation}
\DeclareMathOperator*{\argmin}{arg\,min}
\DeclareMathOperator*{\argmax}{arg\,max}

\usetikzlibrary{plotmarks,automata,positioning}
\usetikzlibrary{decorations.markings}
\tikzstyle{axis}   = [-latex,black!55]
\tikzstyle{two}=[x={(1cm,0cm)},y={(0cm,1cm)}]
\tikzstyle{dim}    = [latex-latex]

\setlist[itemize]{leftmargin=*}

\begin{document}

\title{Truthful Facility Location with Additive Errors}

\author{
	Iddan Golomb\thanks{Tel Aviv University. Email: {\tt igolomb@gmail.com}.
		The work of Iddan Golomb was partially supported by the European Research Council under the European Union's Seventh Framework Programme (FP7/2007-2013) / ERC grant agreement number 337122. This work was done in part while I. Golomb was visiting the Simons Institute for the Theory of Computing.} \and
	Christos Tzamos \thanks{Massachusetts Institute of Technology. Email: {\tt tzamos@mit.edu}.
		This work was done in part while C. Tzamos was visiting the Simons Institute for the Theory of Computing.}
	}

\date{January 2, 2017}

\maketitle

\begin{abstract}
We address the problem of locating facilities on the $[0,1]$ interval based on reports from strategic agents. 
The cost of each agent is her distance to the closest facility, and the global objective is to minimize either the maximum cost of an agent or the social cost.

As opposed to the extensive literature on facility location which considers the multiplicative error, we focus on minimizing the worst-case {\em additive} error.
Minimizing the additive error incentivizes mechanisms to adapt to the size of the instance. 
I.e., mechanisms can sacrifice little efficiency in small instances (location profiles in which all agents are relatively close to one another), in order to gain more [absolute] efficiency in large instances.
We argue that this measure is better suited for many manifestations of the facility location problem in various domains.

We present tight bounds for mechanisms locating a single facility in both deterministic and randomized cases. We further provide several extensions for locating multiple facilities.
\end{abstract}

\section{Introduction}
We consider a setting in which several agents are located on a line, and a central planner intends to place a facility at some point upon that line. 
Each of the agents seeks to be as close as possible to the facility -- the cost of an agent is her distance to the facility.
The planner wishes to minimize some global objective -- either the average or the maximum cost of an agent.
In some cases, agents might misreport their preference in order to obtain a better outcome from their perspective.
We aim to devise \textit{truthful} mechanisms, in which no agent can benefit from misreporting, regardless of the reports of the other agents. 

In the literature, the most common measure for assessing a mechanism's efficiency is the approximation ratio -- the worst-case ratio (over any location profile) between the global cost of the mechanism and that of the optimal mechanism. 
In this paper, we strive to minimize the \textit{additive} error -- the worst-case \textit{difference} between the same two values.

To illustrate the difference between these two error functions, consider the classical facility location problem in which a municipality wants to build a library on one of its streets in order to serve the residents of the street. 
Optimizing over the approximation ratio might guarantee, for instance, that no resident will need to walk to the library more than \textit{2 times} the maximal walking distance of an agent in the optimal assignment. In contrast, optimizing over the additive error might guarantee that no agent will need to walk to the library for more than \textit{3 kilometers} over the optimal assignment.
This example demonstrates that optimizing over the additive error guarantees efficiency with respect to a tangible and natural measure of absolute distance, and further provides a target function which is easily understood. 

As opposed to the multiplicative error which is scale-invariant, by focusing on the additive error we sacrifice some efficiency in small instances (e.g., cases in which all residents are close to the library), in which the solutions are already quite good, so as to achieve a nearly optimal solution in larger instances (where the error is significantly higher).
Indeed, in many cases some (large) instances are more important to the designer than other (smaller) instances, and the additive error accommodates for this need.
 
Clearly, when the size of the instance doubles, the additive error also doubles. 
However, in many different manifestations of the facility location problem, the domains are bounded. We restrict ourselves to such cases, i.e., in our setting the network is not $\Re$, but rather the $[0,1]$ segment of the line.
Some examples of such settings include:

\begin{itemize}
	\item In the aforementioned example of locating a public library on a street, there is clearly a realistic bound on the length of a street, e.g., the length of the longest street in the city. 
	\item A group of public officials needs to vote over the amount of resources to invest in a given policy (political, social, etc.).  
	E.g., the Ministry of Education is faced with the decision of how much money the government should allocate to a reform in the school system.
	Note that in this case scaling can be done with respect to the size of the total education budget, since decisions which affect millions of dollars tend to be more important than those which affect thousands of dollars.

	\item A group of people sitting in a room want to collectively decide on the temperature of the air conditioner. 
	Each person has their own ideal temperature, and some reasonable assumptions can be used to bound the minimal and maximal temperature values (e.g., it is probably safe to assume that nobody's true preference is 100 degrees Celsius).
	
	\item A group of colleagues wishes to set up a meeting in the upcoming week, and each one has an ideal time slot for the meeting. 
	In some cases it is assumed that the utility function is single-peaked (see, e.g., \cite{schummer2002strategy}). 
	In this example, time provides a natural scale for the problem and it seems reasonable to get solutions where everybody waits at most 5 minutes more than the optimal, even if the optimal solution is $0$.
	In contrast, if the optimal solution is $3$ hours, a $2$ approximation can be very bad. 
	
\end{itemize}
Assuming quasi-linear utilities and allowing payments, then the well known Vickrey-Clarke-Groves (VCG) mechanism is truthful and can achieve the optimal social cost (e.g., \cite{groves1973incentives}). However, in many real-life situations (such as in the examples previously mentioned) we restrict the use of money due to ethical, legal or other considerations.

Given a set of votes, it is polytime to find the optimal outcome. However, when restricted to truthful mechanisms, we show that the optimal result cannot be selected in the general case. In other words, approximation is used to circumvent truthfulness and not computational hardness.



\subsection{Our Contributions}
We analyze the worst-case additive error between truthful mechanisms and the optimal mechanism for two objectives - the maximum cost and the average cost of an agent. The average cost is the same as the more commonly used objective of the social cost up to a multiplicative factor of $n$ (the amount of agents), and we chose to focus on the average cost purely due to convenience as the results are easier to interpret without the parameter $n$.

It 
In Observation \ref{obs-nplusone}, we show that the error for the average cost grows at least linearly with $n$.
Using this normalization, it is clear that since the network is the $[0,1]$ line, the additive error for both the average cost and the maximum cost of any mechanism is always between $0$ and $1$. 
For example, the trivial mechanism which locates the facility at $1/2$ regardless of the agents' reports has an additive error of $1/2$ for both the social and maximum cost.
Our choice of an interval length of $1$ is only done for the sake on convenience, and our results hold without loss of generality for any interval $[0,M]$ with an obvious multiplicative adjustment of the ratios (e.g.,	 $M/6$ instead of $1/6$). 

In the paper, we show the following results:
\begin{itemize}
	\item Locating a single facility (Figure \ref{fig:SingleFacilityResults}) - We provide tight bounds for all variations of this problem. 
	For the average cost, it is well known that locating the facility on the median report is both truthful and optimal. 
	For the maximum cost, we start by presenting a randomized lower bound of $1/6$.
	In order to do so, we characterize the structure of the optimal mechanism, and show that it must locate the facility at one of five possible points (the leftmost and rightmost reports, the center between them, $0$ and $1$). 
	We further show that it must be symmetric (a notion we formally define in the paper).
	We then present two location profiles, and utilize this characterization to exhibit the lower bound.
	We move on to devise a mechanism called BLRC, which is a probability distribution over two known mechanisms, that matches the lower bound of $1/6$. 
	In contrast, the best upper bound of currently known mechanisms was $1/4$ (LRC: left-right-center by \cite{procaccia2009approximate}). 

	For deterministic mechanisms, we show a lower bound of $1/4$. 
	We then present a simple deterministic mechanism called phantom-half that matches this bound. In contrast, the best upper bound of any dictatorship is $1/2$.
	
	\begin{figure}[ht]	
		\begin{framed}

			\caption{Our results for single facility problem with the maximum cost objective. 
				We show tight bounds of $1/6$ for randomized mechanisms and $1/4$ for deterministic mechanisms. 
				The best results for currently known mechanisms were $1/4$ and $1/2$ for randomized and deterministic mechanisms respectively.}
			\centering
			\begin{tabular}{|c|c|c|}
				\hline
				& Randomized & Deterministic \\ \hline
				Lower & $1/6$  & $1/4$ \\ 
				Bound & Thm. \ref{thm:RanLB1Fac} & Thm. \ref{thm:DetLB1Fac} \\    \hline
				Upper & $1/6$ &  $1/4$ \\ 
				Bound & BLRC, Thm. \ref{thm:RanUB1Fac}  &  Phantom-half, Thm. \ref{thm:DetUB1Fac} \\ \hline
				Known &  $1/4$ & $1/2$ \\
				Mechanisms & LRC (from \cite{procaccia2009approximate}) &  Any dictatorship\\ 		 	
				\hline		 	
			\end{tabular} 
			\label{fig:SingleFacilityResults}
		\end{framed}	
	\end{figure}

	\item Extensions to multiple facilities - We first show a trivial mechanism, Equal Spread, which locates $k$ facilities with equal distances from one another (without using the instance data) and has a maximum cost of $\frac{1}{2k-1}$. 
	We then extend the construction of the deterministic lower bound for one facility to the multiple-facility setting, and reach a lower bound of $\frac{1}{6k}$, which shows that the result for this problem is $\Theta(1/k)$.

	For the average cost, we present two extensions to Equal Spread which slightly improve the error - the randomized paired-equal-cost (PEC) which has an error of $\frac{1}{4k-2}$ (Theorem \ref{thm:pec}) and the deterministic election-parity-equal-cost (EPEC) mechanism which reaches an error of $\frac{3}{8k-4}$ (Theorem \ref{thm:epec}). 
	In addition, for the special case of 2 facilities, we show a deterministic mechanism based on percentiles which achieves an upper bound of $1/5$ (compared to $1/4$ by EPEC).

\end{itemize}

\subsection{Related Work}

The truthful facility location problem has rich roots.
The initial focus was mainly on characterizing the class of truthful mechanisms - in 1980 Moulin characterized all deterministic truthful mechanisms for locating a facility on the line, when the preferences are single-peaked \cite{moulin1980strategy}, and this characterization was later extended to general graphs by Schummer and Vohra \cite{schummer2002strategy}.

Procaccia and Tennenholtz were the first to prove bounds on the approximation ratio of the game-theoretic facility location problem \cite{procaccia2009approximate}.
Their initial model has been extended by these authors and by others in many ways, and most of these extensions leads to additional bounds on the approximation ratio ---
The network was extended from the line to cycles (\cite{alon2009strategyproof}, \cite{alon2010walking}), trees (\cite{alon2009strategyproof}, \cite{feldman2013strategyproof}) and general graphs \cite{alon2009strategyproof};
Some papers consider building several facilities, where the cost of an agent is her distance to the closest facility (e.g. \cite{lu2009tighter}, \cite{lu2010asymptotically}, \cite{fotakis2013strategyproof}, \cite{fotakis2014power});
Other papers look at heterogeneous facilities, i.e., facilities serving different purposes \cite{serafino2015truthful};
Several papers consider a setting in which every agent possesses multiple locations and pays the sum of the distances to her locations (\cite{procaccia2009approximate}, \cite{lu2009tighter}); 
Additional objective functions were considered besides the maximum cost and the social cost, for instance the $L_2$ norm \cite{feldman2013strategyproof} or the minimax envy \cite{caifacility} (in the latter, the approximation was done with an additive error);
Additional papers consider different preferences of the agents, for instance doubly-peaked preferences \cite{filos2015facility}, ``obnoxious facility location" in which agents want to be as far away as possible from the facility \cite{cheng2013strategy}, settings which combine agents with ordinary preferences and agents who wish to be far from the facility (\cite{feigenbaum2015strategyproof}, \cite{zou2015facility});
Another direction that was researched is the tradeoff between the approximation ratio and the variance (\cite{procacciaapproximation});
Some papers consider different methods of voting, for instance by restricting the outcome to a discrete set of candidates (\cite{dokow2012mechanism}, \cite{sui2015approximately}, \cite{feldman2015voting}) or by using mediators \cite{babaioff2016mechanism}.
When restricting the location of the facility to given candidates, minimizing the social cost of facility location problems has been associated to the notion of ``distortion" (\cite{procaccia2006distortion}, \cite{anshelevich2015approximating}, \cite{feldman2015voting}, \cite{anshelevich2015randomized}, \cite{gross2017vote}).

Nissim et. al. also use additive errors for facility location, as an example of their framework which maintains differential privacy \cite{nissim2012approximately}. However, their model differs from our setting as they grant the mechanism the additional power to impose agents to connect to a specific facility. Imposition allows for better approximation ratios, as seen for instance in \cite{fotakis2010winner}.



\section{Model}
Let $N=\{1, \ldots, n\}$ be a set of agents, where each agent $i \in N$ is located at some point $x_i \in [0,1]$.
The vector $\bx=(x_1, \ldots, x_n)$ is known as a location profile.
A {\em deterministic mechanism} for locating $k$ facilities is a mapping from some location profile (the reports of the agents) to a set of $k$ points on the interval (the chosen locations of the facilities), that is: $M: [0,1]^n \rightarrow [0,1]^k$.

Assuming the facility locations are $M(\bx)=\{l_1,\ldots,l_k\}$, the {\em cost} of point $p$ is its distance to the closest facility: $\cost_p(M,\bx) = \min_{1\leq j \leq k}|l_j-p|$. 
The cost of agent $i$ located at $x_i$ is defined as the cost of her location, $\cost_{x_i}(M,\bx)$.
Each agent aims to minimize her cost.

A {\em randomized mechanism} is a mapping from a location profile to some distribution over $k$-tuples of locations: $M: [0,1]^n \rightarrow \Delta([0,1]^k)$.
The cost of agent $i$ is the expected cost of this agent according to the probability distribution returned by the mechanism, that is: $\cost_{x_i}(M, \bx)=\mathbb{E}_{(l_1,\ldots, l_k) \sim M(\bx)}\min_j|x_i-l_j|$.

A {\em truthful} mechanism $M$ (also known as a \textit{strategyproof} mechanism) is one in which an arbitrary agent cannot suffer from reporting her real location, regardless of the reports of the other agents: For all $i \in N$, all $x_i, x_i' \in [0,1]$ and $\bx_{-i} \in [0,1]^{n-1}$: 
$$cost_{x_i}(M,(x_i,\bx_{-i})) \leq cost_{x_i}(M,(x_i',\bx_{-i})).$$
For randomized mechanisms, these mechanisms are often denoted by the term {\em truthful in expectation }mechanisms, in order to distinguish them from universally truthful mechanisms (a stronger notion which we do not use in this paper, in which an agent cannot regret reporting her true location ex-post).

\sloppy For a location profile $\bx$ and reported locations $\bx'$ the {\em average cost} is: $AC(M,\bx, \bx') = \allowbreak \frac{1}{n} \sum_i \cost_{x_i}(M, \bx')$. 
Note that we scale the sum of the agents' costs by a factor of $n$.
For truthful mechanisms, we can drop the misreport in the notation and denote the average cost by $AC(M,\bx)$.
Given a location profile $\bx$, the optimal average cost is: $AC(OPT,\bx)=\min_{(l_1 \ldots l_k) \in [0,1]^k} \frac{1}{n} \sum_i \min_j|l_j-x_i|$.
For a truthful mechanism $M$, the {\em additive error given a location profile $\bx$} is the difference between the average cost of $M$ and the average cost of the optimal mechanism: $\sigma_M(\bx) = AC(M,\bx) - AC(OPT,\bx)$.
The {\em additive error} of a truthful mechanism $M$ is the maximal error over any location profile: $\sigma_M = \max_{\bx} \sigma_M(\bx)$.

Given location profile $\bx$ and reported locations $\bx'$ the {\em maximum cost} (max-cost in short) is: $MC(M,\bx, \bx') = \max_i \cost_{x_i}(M, \bx')$. 
Similarly to the average cost case, for a truthful mechanism $M$ we denote the max-cost by $MC(M,\bx)$. 
Given a location profile $\bx$, the optimal maximum cost is: $$MC(OPT,\bx)=\min_{(l_1 \ldots l_k) \in [0,1]^k}\max_i \min_j|l_j-x_i|$$
For a truthful mechanism $M$, the {\em additive error given a location profile $\bx$} is the difference between the maximum cost of $M$ and the maximal cost of the optimal mechanism: $\delta_M(\bx) = MC(M,\bx) - MC(OPT,\bx)$.
The {\em additive error} of a truthful mechanism $M$ is the maximal error over any location profile: $\delta_M = \max_{\bx} \delta_M(\bx)$.

Given a location profile $\bx$, we denote the leftmost and rightmost points in $\bx$ by $\xl$ and $\xr$ respectively. The center point between these two points is called $\xc = \frac{\xl+\xr}{2}$.

We now show that lower bounds proven for few agents can also be extended to location profiles with arbitrarily many agents since the error grows at least linearly with the amount of agents.
The first observation shows that any lower bound for both the maximum and average cost proven with a location profile with $n$ agents also holds for any profile with $q\cdot n$ agents for an arbitrary positive integer $q$ (i.e., for any profile with an even number of agents).

\begin{observation} \label{obs-amountOfAgents}
	Let $M$ be a truthful mechanism which has an additive error of $\epsilon$ for any profile with $q \cdot n$ agents for some global cost function (either the maximum or the average cost) with $q \geq 1$.
	Then there exists some truthful mechanism $M'$ which has an additive error of $\epsilon$ for the same global cost function, for any profile with $n$ agents.
\end{observation}

\begin{proof}
	Let $M'$ be the mechanism which simulates $M$ based on the location profile which duplicates each report $q$ times. 
	That is, for some input $\bx_A=(x_1^A,\ldots,x_n^A)$, $M'$ creates the profile

				   
	$\bx_B=( x_1^B=x_2^B=\cdots=x_q^B=x_1^A,x_{q+1}^B=x_{q+2}^B=\cdots=x_{2q}^B=x_2^A,\allowbreak \cdots, \allowbreak x_{(n-1)q+1}^B=x_{(n-1)q+2}^B=\cdots=x_{nq}^B=x_n^A )$ and locates the facilities on $M'(\bx_A)=M(\bx_B)$.
	
In these profiles in which the profile is duplicated, the additive error of these mechanisms is the same (and therefore in the general case, the cost of $M$ is at least as large as that of $M'$): for the maximum cost it holds that:
\begin{align*}MC(M',\bx_A)&=\max_{i}|x_i^A-M'(\bx_A)|\\&=\max_{j}|x_j^B-M(\bx_B)|=MC(M,\bx_B).\end{align*}
	For the average cost, both the numerator and the denominator are multiplied by $q$: 

\begin{align*}AC(M',\bx_A)&=\frac{1}{n}\sum_{i=1}^n |x_i^A-M'(\bx_A)|\\&= \frac{1}{qn}\sum_{j=1}^{qn} |x_j^B-M(\bx_B)| =MC(M,\bx_B).\end{align*}

	Clearly, the cost of the optimal mechanism is the same for both of these cases as well.
	
	It is only left to prove truthfulness under $M'$. This holds due to the lemma by Lu et al. in \cite{lu2010asymptotically} in which they show that any strategyproof (that is, \textit{truthful} according to our terminology) mechanism is also partial group strategyproof. 
	A mechanism is defined to be partial group strategyproof if for any group of agents on the same location, each individual cannot benefit if they misreport simultaneously.
	Formally, given a non-empty set $S \subset N$, profile $\bx=(\bx_S,\bx_{-S})$ where $\bx_S=(x,\ldots,x)$ for some $x$, and some misreported locations $\bx_S'$ it holds that for any $i \in S$: $$\cost_i(M(\bx_S,\bx_{-S}) \leq \cost_i(M(\bx_S',\bx_{-S})).$$ If there had been some beneficial deviation of some $x_i$ to $x_i'$ in $M'$ then there must have been some deviation of the coalition $(x_i,x_i)$ to $(x_i',x_i')$ in $M$, in contradiction to partial group strategyproofness.
\end{proof}
The next observation shows that any lower bound the maximum cost proven with a location profile with $n$ agents also holds for any profile with $n+1$ agents. Clearly this observation can be applied repeatedly to show a lower bound for an arbitrary amount of agents.

\begin{observation} \label{obs-nplusone}
	Let $M$ be a truthful mechanism which has an additive error of $\epsilon$ for any profile with $n+1$ agents for the maximum cost. Then there exists some truthful mechanism $M'$ which has an additive error of $\epsilon$ for the the maximum, for any profile with $n$ agents.
\end{observation}

\begin{proof}
	Let $M'$ be the mechanism which simulates $M$ based on the location profile for which the last agent is placed together the penultimate agent. 	That is, for some input $\bx_A=(x_1^A,\ldots,x_n^A)$, $M'$ creates the profile $\bx_B=\left(x_1^B=x_1^A, x_2^B=x_2^A, \ldots, x_n^B=x_n^A, x_{n+1}^B=x_n^A \right)$ and locates the facilities on $M'(\bx_A)=M(\bx_B)$.
	
	Proof of truthfulness and the additive error follows the same lines as the previous observation -- Truthfulness holds due to to partial group strategy proofness in $M$. The maximum error of $M$ and $M'$ is the same since the outputted location is the same ($M'(\bx_A)=M(\bx_B)$) and the set of the locations in the inputs (the distinct locations in $\bx_A,\bx_B$) are the same.
\end{proof}

\section{Locating a Single Facility}
\subsection{Randomized Mechanisms}

For the single facility problem, Procaccia and Tennenholtz introduced the LRC (left-right-center) mechanism, which chooses $\xl$ and $\xr$ with probability $\frac{1}{4}$ each, and chooses $\xc$ with probability $\frac{1}{2}$ \cite{procaccia2009approximate}. 
They further showed that LRC is truthful-in-expectation and achieves a tight bound for the multiplicative ratio. 
It is easy to see that LRC achieves an additive error of $\frac{1}{4}$ (for the location profile $\bx=(0,1)$).

We start by showing a randomized lower bound of $\frac{1}{6}$. We match this bound with a mechanism which extends LRC (called Balanced-LRC or BLRC in short). 

The following two definitions set the foundations for the proof of the lower bound.

\begin{definition}[5-point mechanism]
	For an arbitrary location profile $\bx$, a 5-point mechanism $M$ is one which can only assign a positive probability to a subset of the following 5 points: $0,x_L,x_C,x_R,1$. 

\end{definition}

We denote the probabilities that $M$ locates the facilities on these points as $p_0,p_L,p_C,p_R,p_1$ respectively.

\begin{definition}[Symmetric 5-point mechanism]
~\\A $5$-point mechanism $M$ is termed \textit{symmetric} if for any location profile $\mathbf{x}$ such that $\mathbf{x}=\mathbf{1}-\mathbf{x}$ it holds that $p_L=p_R$ and $p_0=p_1$.
	
\end{definition}

\begin{theorem} \label{thm:RanLB1Fac}
	Any randomized truthful in expectation mechanism for locating one facility has an additive error of at least $\frac{1}{6}$ for the maximum cost.
\end{theorem}
\begin{proof}
	We present a lower bound for two agents.
	First, we show that any truthful in expectation mechanism for 2 agents can be replaced with a 5-point truthful in expectation mechanism while conserving the additive error. 
	Then, we show that we can further restrict ourselves to \textit{symmetric} 5-point mechanism without increasing the additive error.
	This characterization is used by a pair of location profiles and a transition between them, yielding the lower bound.
	Finally, due to Observation \ref{obs-nplusone}, this lower bound can be extended to an arbitrary amount of agents - for any number of agents $n$ there does not exist a mechanism with a lower additive error than $\frac{1}{6}$.	
	\begin{lemma}
		If there exists a truthful in expectation mechanism $M$ which has additive error $\alpha$ for the maximum cost for 2 agents, then there exists a truthful in expectation 5-point mechanism $M'$ which also has additive error $\alpha$ for the maximum cost for 2 agents.
	\end{lemma}
	\begin{proof}
		The proof will transform $M$ to $M'$ by moving the probability that $M$ allocated to any point $z$, to the left and right neighbors of $z$ from the set $A= \{0,x_{L},x_C,x_{R},1\}$, without changing the expected location. 
		The resulting mechanism will clearly be a 5-point mechanism, and the following claims will show that it also preserves truthfulness and the additive error.
		We start by showing the effect of this transformation on an arbitrary deterministic mechanism $M_1$.
		
		For some arbitrary point $z$, let $z_1,z_2$ be its left and right neighbors, respectively, from the set $A$, that is: 
		\begin{eqnarray*}
			z_1 = \argmax_{x \in A} (x \leq z) \\
			z_2 = \argmin_{x \in A} (x \geq z) 
		\end{eqnarray*}			
		Let $Z$ be the random variable which can take two possible values, $z_1$ and $z_2$, such that $\mathbb{E}[Z]=z$ (that is $Pr(M_2(\mathbf{x})=z_1)=\frac{z_2-z}{z_2-z_1},Pr(M_2(\mathbf{x})=z_2)=\frac{z-z_1}{z_2-z_1}$).
		For some location profile $\mathbf{x}$, let $M_1$ be the deterministic mechanism which locates the facility at point $z$, and let $M_2$ be the randomized mechanism which locates the facility according to the random variable $Z$.
		
		\begin{claim} \label{clm:outsideRegion}
			For any $p \in [0,z_1] \cup [z_2,1]$, the cost of $p$ under $M_1$ is the same as its cost under $M_2$, that is: 
			$$\cost_{p}(M_1,\mathbf{x})=\cost_{p}(M_2,\mathbf{x}).$$
		\end{claim}
		\begin{proof}
			
		Assume without loss of generality that $p \leq z_1 < z_2$.
		The cost of $p$ under $M_1$ is: $\cost_{p}(M_1,\mathbf{x})=|p-M_1(x)|=z-p$. 
		
		On the other hand, the cost under $M_2$ is:
		\begin{align*}
			\cost_{p}(M_2,\mathbf{x}) &=\mathbb{E}|Z-p| = \frac{z_2-z}{z_2-z_1}(z_1-p)+ \frac{z-z_1}{z_2-z_1}(z_2-p) \\
			&= \frac{z_2-z}{z_2-z_1}(z_1-p)+ \frac{z-z_1}{z_2-z_1}[(z_2-z_1)+(z_1-p)] \\
			&= (z_1-p)+\frac{z-z_1}{z_2-z_1}(z_2-z_1) = z-p 
		\end{align*}
    Thus, both costs are the same.
		\end{proof}
		
		\begin{claim} \label{clm:insideRegion}
			For any $p \in [0,1]$, the cost of $p$ under $M_1$ is less than or equal to its cost under $M_2$: 
			$\cost_{p}(M_1,\mathbf{x}) \leq \cost_{p}(M_2,\mathbf{x})$.
		\end{claim}
		\begin{proof}
			The cost under $M_1$ is clearly $\cost_{p}(M_1,\mathbf{x})=|p-z|$.
			
			For $M_2$ it holds that:
			\begin{eqnarray*}
				\cost_{p}(M_2,\mathbf{x}) = \mathbb{E}|Z-p| \geq |\mathbb{E}[p-Z]| = |p - \mathbb{E}[Z]| = |p-z| 
			\end{eqnarray*}			
			This holds due to Jensen's inequality (for any convex function $f$ and random variable $Z$: $f(\mathbb{E}[Z]) \leq \mathbb{E}(f(Z))$), since the absolute value is a convex function.
		\end{proof}
		
		These two claims and truthfulness of $M_1$ collectively show that if $M_1$ was truthful, then so is $M_2$: let $\bx$ be a location profile, and let $\bx'$ be the profile after deviation of some $x_i \in \{x_L,x_R\}$ to $x_i'$. Then: $\cost_{x_i}(M_2,\bx)=\cost_{x_i}(M_1,\bx) \leq \cost_{x_i'}(M_1,\bx)\leq \cost_{x_i'}(M_2,\bx)$.

		\begin{claim}
			The maximum cost of $\mathbf{x}$ under $M_1$ is equal to the maximum cost under $M_2$.
		\end{claim}
		\begin{proof}
			Assume without loss of generality that $z \leq x_C$. Therefore, $x_R$ incurs the maximum cost under $M_1$, and this cost is $x_R-z$. Due to Claim \ref{clm:outsideRegion}, this is also the cost of $x_R$ under $M_2$. 
			Since in $M_2$ the probability is split only between points $x_L,x_C$ (if $x_L < z \leq x_C$) or points $0,x_L$ (if $0 \leq z \leq x_L$), then clearly the agent with the maximum cost in $M_2$ is $x_R$.
		\end{proof}
		
		To conclude the proof of the lemma, it is only necessary to repeat the above process for any point $z \notin A$ which is chosen with a positive probability.
	\end{proof}

	We now show that we can restrict ourselves to \textit{symmetric }5 point mechanisms.
	
	\begin{lemma}
		For any truthful in expectation 5-point mechanism $M_1$ which has an additive error $\alpha$, there exists a truthful in expectation symmetric 5-point mechanism $M_2$ which also has an additive error of $\alpha$.
	\end{lemma}
	
	\begin{proof}
		Let $M_1$ be an arbitrary truthful 5-point mechanism which achieves an additive error $\alpha$. 
		We will define a mechanism $M_3$, and use it to construct a truthful in expectation symmetric mechanism $M_2$.
		
		We start by defining $M_3$. 
		For an arbitrary profile $\bx$ denote the probabilities $M_1$ chooses for $0,x_{L},x_C,x_{R},1$ as $p_0^1,p_L^1,p_C^1,p_R^1,p_1^1$ respectively. 
		Let $M_3$ be a 5-point mechanism which does the following: For an arbitrary profile $1-\bx$, $M_3$ locates the facility based on points $0,1-x_R,1-x_C,1-x_L,1$ with the following probabilities: $p_0^3=p_1^1$, $p_L^3=p_R^1$, $p_C^3=p_C^1$, $p_R^3=p_L^1$ and $p_1^3=p_0^1$ (notice that the leftmost and rightmost points in $1-\bx$ are $1-x_R$ and $1-x_L$ respectively). 
		In some sense $M_3$ can be seen as the ``anti-symmetric" mechanism of $M_1$.

		$M_3$ is truthful -- Assume towards a contradiction otherwise, that is, there exists some $\bx$ and a beneficial deviation to $\bx'$. Then, the deviation from $1-\bx$ to $1-\bx'$ would have been beneficial in $M_1$, in contradiction to the fact that $M_1$ is truthful. 
		Additionally, following the same logic, the maximum cost of $M_3$ is equal to that of $M_1$ (for any profile $\bx$ in $M_3$ with error $\alpha$, the profile $1 - \bx$ has an error of $\alpha$ in $M_1$).	
		
		Let $M_2$ be the mechanism which runs $M_1$ and $M_3$ with probability $0.5$ each. $M_2$ is a probability distribution over truthful mechanisms, therefore it is necessarily truthful. 
		Also, the maximum cost is the same in $M_1,M_3$, and it is therefore also the maximal cost in $M_2$.
		Finally, by the construction of $M_2$ it is clearly a symmetric 5-point mechanism.
	\end{proof}
	
	Now that we have proven that we can focus only on symmetric 5-point mechanisms, we present two location profiles and a transition between them which concludes the proof.
	
	Let $\mathbf{x_A}=(\frac{1}{3},\frac{2}{3})$ and $\mathbf{x_B}=(0,\frac{2}{3})$ be two location profiles. 
	Let $M$ be a symmetric 5-point mechanism which achieves the optimal additive error. 
	
	For profile $\mathbf{x_A}$, let the probability of locating the facility on the left (that is $1/3$) be $p_{LR}^A$ and the probability of locating the facility in the center point (that is $1/2$) be $p_C^A$. Since the instance is symmetric, the probability of locating the facility on the right point ($2/3$) will also be $p_{LR}^A$ and the probability of locating the facility at $0$ and likewise at $1$ is $p_{01}^A=1/2-p_{LR}^A-\frac{p_C^A}{2}$.
	
  We have that the cost of the agent located at the left is equal to:
	\begin{align*}	
	\cost_{1/3}(A) &= p_{01}^A \left(\frac{1}{3}+\frac{2}{3} \right) + \frac{ p_{LR}^A}{3} +  \frac{p_C^A}{6} \\
				&=	p_{01}^A  + \frac{ p_{LR}^A}{3} +  \frac{p_C^A}{6} \\
				&= 1/2 - p_{LR}^A - p_C^A/2 + p_{LR}^A /3 + p_C^A / 6 \\
				&=  \frac{1}{2} - \frac{p_C^A}{3} - \frac{2p_{LR}^A}{3} 
	\end{align*}	
  We now consider the profile $B$. The additive error of the profile is equal to 
  $$\mathbb{E}[ \max\{|X|,|X-\frac 2 3|\}] - \frac 1 3 = \mathbb{E} |X-\frac 1 3| =\cost_{1/3}(B). $$
	However, since the mechanism is truthful it holds that $$\cost_{1/3}(A) \leq \cost_{1/3}(B).$$ 
	If we assume towards a contradiction that  $M$ can achieve an additive error of less than $\frac{1}{6}$, we have that 
  $\cost_{1/3}(B)<\frac 1 6$ and consequently $\cost_{1/3}(A) < 1/6$.
	This implies that $\frac{1}{2} - \frac{p_C^A}{3} - \frac{2p_{LR}^A}{3} < \frac 1 6$ which gives $p_C^A + 2p_{LR}^A > 1$.
	This is contradiction since in profile $A$ probabilities must sum to $1$.
\end{proof}

We now present a new mechanism, and show that it can achieve a matching upper bound of $1/6$: 
\begin{definition}[BLRC mechanism]~\\
	The balanced left-right-center (BLRC) mechanism locates the facility at the point $\frac{1}{2}$ with probability $\frac{1}{3}$, and deploys the LRC mechanism with probability $\frac{2}{3}$.
	More explicitly, BLRC locates the facility according to the following distribution: 
	$$
  \frac{1}{2}, \text{ w.p. }  \frac{1}{3}; \quad
	x_{L} \text{ w.p. }  \frac{1}{6}; \quad
	x_{R} \text{ w.p. }  \frac{1}{6}; \quad
	x_C \text{ w.p. }  \frac{1}{3}
	$$
\end{definition}

\begin{theorem} \label{thm:RanUB1Fac}
	BLRC is truthful in expectation, and achieves an additive error of at most $\frac{1}{6}$ for the maximum cost.
\end{theorem}
\begin{proof}
	BLRC is based on two truthful-in-expectation mechanisms -- LRC was proven to be truthful in expectation in \cite{procaccia2009approximate}, and locating the facility on the fixed point $1/2$ is clearly truthful as the agents have no influence over the result.
	Therefore taking a distribution over these two mechanisms is also truthful in expectation.
	
	Proof of the approximation ratio will be based on two lemmas:
	\begin{lemma} \label{lma:moveRight}
		Let $x,y$ be points such that: $0 \leq x < y \leq 1$, and let $\bx,\bx'$ be the following two location profiles: $\bx=(x,y)$ and $\bx'=(0,y-x)$. Then $\delta_{BLRC}(\bx) \leq \delta_{BLRC}(\bx')$.
	\end{lemma}
	\begin{proof}
		For the profile $\bx$, the optimal mechanism locates the facility at $\frac{x+y}{2}$ and its cost is $\frac{y-x}{2}$.
		The maximum cost of $BLRC$ in this case is: $\frac{1}{3}\cdot(y-x)+ \frac{1}{3}\frac{y-x}{2}+\frac{1}{3}\cdot(\max\{|y-1/2|,|x-1/2|\})$.
		Therefore the additive error is: $\delta_{BLRC}(\bx)= \frac{\max\{|y-1/2|,|x-1/2|\}}{3}$.
		
		For the profile $\bx'$ the cost of the optimal mechanism remains $\frac{y-x}{2}$, but the cost of $BLRC$ is: $\frac{1}{3}(y-x)+ \frac{1}{3}\cdot \frac{y-x}{2}+\frac{1}{3}\cdot \frac{1}{2}$. 
		Therefore: $\delta_{BLRC}(\bx') = 1/6$.		
		
		For any $x,y \in [0,1]$ it holds that $\frac{1}{2} \geq \max\{|y-1/2|,|x-1/2|\}$, therefore:
		$\delta_{BLRC}(\bx') - \delta_{BLRC}(\bx) = \frac{1}{3}(\frac{1}{2}-\max\{|y-1/2|,|x-1/2|\}) \geq 0$.
	\end{proof}
	\begin{lemma} \label{lma:maxErrorSixth}
		For any $z \in [0,1]$, let $\bx = (0,z)$. It holds that the additive error of BLRC for $\bx$ is $1/6$: $\delta_{BLRC}(\bx)=\frac{1}{6}$.
	\end{lemma}
	\begin{proof}
		Given the location profile $\bx$, the optimal mechanism locates the facility at $z/2$ at a maximum cost of $z/2$.
		The cost of $BLRC$ is: $MC(BLRC,\bx)=\frac{1}{3}(z)+ \frac{1}{3}\cdot \frac{z}{2}+\frac{1}{3}\cdot \frac{1}{2}$. 
		
		Therefore, the additive error is: $\delta_{BLRC}(\bx) = \frac{1}{3}\cdot\frac{z}{2}+\frac{1}{3}\cdot(\frac{1}{2}-\frac{z}{2})=\frac{1}{3}\cdot\frac{1}{2}=\frac{1}{6}$.
	\end{proof}
	
	Lemma \ref{lma:moveRight} holds for an arbitrary location profile $\bx$ with $x_{L}=x,x_{R}=y$, and Lemma \ref{lma:maxErrorSixth} holds for an arbitrary $z$, and in particular $z=y-x$, so the error of BLRC with two agents is no more than $1/6$. Since the maximum cost always occurs in at least one of the two extreme reports, combining the two lemmata is sufficient to complete the proof for an arbitrary number of agents as well.
  $\blacksquare$
\end{proof}

\subsection{Deterministic Mechanisms}
We move on to prove a lower bound of $1/4$ and a matching upper bound.
Recognize that any dictatorship has an additive error of $1/2$, by considering $\bx=(0,1)$.

\begin{theorem} \label{thm:DetLB1Fac}
	Any deterministic truthful mechanism $M$ has an additive error of at least $\frac{1}{4}$ for the maximum cost.
\end{theorem}
\begin{proof}
	We first deal with the case where there are two agents, and later extend to an arbitrary number of agents.
	
	Let $\bx=(0,1)$. Assume without loss of generality that $M(\bx)=1/2+\epsilon$ for some $\epsilon \geq 0$. 
	Let $\bx'=(0,1/2+\epsilon)$. 
	The optimum for $\bx'$ is achieved at $1/4+\epsilon/2$.
	Therefore, if the mechanism is to achieve an additive error of less than $1/4$, it must locate the facility in $(0,1/2+\epsilon/2)$.
	But if this were the case, then the agent at $1/2+\epsilon$ could benefit by misreporting to $1$, which would move the facility precisely to $1/2+\epsilon$, contradicting truthfulness.
	
	For the case of more than two agents, consider the profile in which all other agents are precisely at $1/2$, and repeat the process above. 
\end{proof}

This lower bound construction works in similar lines to the lower bound of Theorem 2.2 in \cite{procaccia2009approximate}.

We now present a matching upper bound, a truthful mechanism additive error of $\frac{1}{4}$.
\begin{definition}[Phantom-half]
	For any profile $\bx$ where $x_L,x_R$ are the leftmost and rightmost locations, $M$ locates the facility on the median of the following 3 points: $x_L,x_R,0.5$.
\end{definition}
The name of this mechanism is inspired by the notion of phantom-voters (in this case - the phantom voter is $0.5$) as introduced in \cite{moulin1980strategy}.

\begin{theorem} \label{thm:DetUB1Fac}
	Phantom-half is truthful and has an additive error of $\frac{1}{4}$ for the maximum cost.
\end{theorem}
\begin{proof}
	It is easy to see that by misreporting, any agent can either not affect the location of the facility, or move it farther away from them.
	
	For the additive error: 
	\begin{itemize}
		\item If $x_L,x_R \leq 1/2$: The optimal mechanism locates the facility at $\frac{x_L+x_R}{2}$ for a maximum cost of $\frac{x_R-x_L}{2}$. Phantom-half locates the facility at $x_R$ for a cost of $x_R-x_L$. The additive error is: $\frac{x_R-x_L}{2} \leq \frac{1/2}{2}=\frac{1}{4}$.
		\item If $x_L\leq 1/2,x_R\geq 1/2$: Assume without loss of generality that $1/2-x_L \geq x_R - 1/2$. The optimal mechanism locates the facility at $\frac{x_L+x_R}{2}$ for a maximum cost of $\frac{x_R-x_L}{2}$. Phantom-half locates the facility at $1/2$ for a cost of $1/2-x_L$. The additive error is: $(1/2-x_L)-\frac{x_R-x_L}{2}=\frac{1-x_L-x_R}{2}$. Since $x_R \geq 1/2,x_L \geq 0$ it holds that: $\frac{1-x_L-x_R}{2} \leq \frac{1}{4}$.
		\item If $x_L,x_R > 1/2$: This is completely symmetric to the first case.
	\end{itemize}
\end{proof}

\section{Extension to Many Facilities}
We start by showing a trivial deterministic mechanism, Equal Spread, which achieves an error of no more than $\frac{1}{2k-1}$ for both the average cost and the maximum cost. We then show a lower bound of $\frac{1}{6k}$ for the maximum cost, meaning the bound is $\Theta(\frac{1}{k})$. 

Afterwards we show two extensions of Equal Spread - PEC, a randomized mechanism which reaches an error of $\frac{1}{4k-2}$ for both max-cost and average cost, and EPEC, a deterministic mechanism which reaches an error of $\frac{3}{8k-4}$ for the average cost.

Finally, we show that for the case of 2 facilities, we can improve the bound of EPEC for the average cost from $1/4$ to $1/5$ by choosing the ``fifths" mechanism, which locates the facilities on the 0.2 and 0.8 percentiles.

\begin{definition}[Equal Spread mechanism]
The	Equal Spread mechanism locates the $k$ facilities on $\frac{i}{2k-1}$ for odd values of $i$ (such that $1 \leq i \leq 2k-1$).
\end{definition}

\begin{theorem}
	Equal spread is a deterministic mechanism which achieves an additive error of no more than $\frac{1}{2k-1}$ for both the average cost and the maximum cost.
\end{theorem}
\begin{proof}
	Trivial - Clearly, for any point $p \in [0,1]$, $p$ is located at a distance of at most $\frac{1}{2k-1}$ to the closest facility. 
\end{proof}


\begin{theorem} \label{thm:DetLBkFac}
	Any deterministic truthful mechanism $M$ for locating $k$ facilities has an additive error of at least $\frac{1}{6k}$ for the maximum cost.
\end{theorem}
\begin{proof}
	The proof separates $k-1$ agents far away from one another, and then uses the proof of Theorem \ref{thm:DetLB1Fac}, except on a smaller interval.
	
	Let $\bx = (x_j)_{j=1}^{k+1}$ be the following location profile: let $x_1=0$, $x_2=\frac{2}{3k}$ and for every $3 \leq i \leq k+1$: let $x_i = \frac{i-1}{k}$. 
	An optimal mechanism can locate $k-1$ facilities on $\{x_3, x_4, \ldots x_{k+1}\}$, and locate one facility on the midpoint of $x_1,x_2$ (that is, on $\frac{1}{3k}$), for a maximum cost of $\frac{1}{3k}$.
	
	Any mechanism with an error lower than $\frac{1}{6k}$ must also designate one facility to the interval in the vicinity of every one of the $k-1$ agents on $x_3 \ldots x_{k+1}$: If there exists some $2 \leq i \leq k-1$ such that there are no facilities in the segment $(\frac{i-1/2}{k},\frac{i+1/2}{k})$ or no facilities on $(\frac{k-1/2}{k},1]$ then there exists an agent located at least $\frac{1/2}{k}$ from a facility, therefore the error would be at least: $\frac{1/2}{k} - \frac{1}{3k} = \frac{1}{6k}$.
		
	Additionally, any mechanism with an error not larger than $\frac{1}{6k}$ must also designate one facility serve the first two agents ($x_1,x_2$): If there are no facilities in the segment $[0,\frac{1}{k}]$ then the error is at least $\frac{1}{k}-0>\frac{1}{6k}$.	
	
	Assume without loss of generality that the remaining facility is put at point $\frac{1}{3k}+\epsilon$ for some $\epsilon \geq 0$.
	Let $\bx'$ be the location profile in for any $j \neq 2$: $x_j'=x_j$ and $x_2'=\frac{1}{3k}+\epsilon$. 
	The aforementioned arguments also dictate that for any $M$ with error less than $\frac{1}{6k}$, in $\bx'$ then $x_1',x_2'$ are also designated one facility (that is, there is a facility in the segment $[0,\frac{1}{k}]$). 
	In profile $\bx'$ the cost of the optimal mechanism is $\frac{1}{6k}+\epsilon/2$. For any $M$ with error less than $\frac{1}{6k}$ the facility must be located in the segment $(\epsilon/2,\frac{1}{3k}+\epsilon/2)$. 
	But if this were the case, then agent $x_2'$ could benefit by misreporting to $\frac{2}{3k}$ (and then the facility would have been located precisely on it), contradicting truthfulness.
%
%
%
	
\end{proof}

We propose a randomized mechanism called PEC which extends this basic idea.

\begin{definition} [Paired-Equal-Cost (PEC) mechanism]
	The paired-equal-cost mechanism locates the $k$ facilities in the following locations, for $0 \leq i \leq 2k-1$:
	\begin{align*}	
		M(\bx) =			
		\begin{cases}
			\{0,\frac{2}{2k-1},\frac{4}{2k-1},\ldots, \frac{2k}{2k-1} \} & \quad \text{ w.p. }  \frac{1}{2} \\
			\{\frac{1}{2k-1},\frac{3}{2k-1},\frac{5}{2k-1},\ldots, 1 \} & \quad \text{ w.p. }  \frac{1}{2} 
		\end{cases}
	\end{align*}
\end{definition}

\begin{theorem} \label{thm:pec}
	PEC is a truthful in expectation randomized mechanism which has an additive error of $\frac{1}{4k-2}$ for the average cost.
\end{theorem}

\begin{proof}
	PEC is clearly truthful since the reports have no effect over the outcome.
	
	We show the additive error by bounding the additive error of an arbitrary agent located at point $x_i$ from above by $\frac{1}{4k-2}$. Let $\frac{j}{2k-1} \leq x_i < \frac{j+1}{2k-1}$ for some $j$ \footnote[1]{It is easy to verify that for $x_i=1$ the cost is $\frac{1}{4k-2}$.}.  	
	Therefore, the closest facility to the agent will be either $\frac{j}{2k-1}$ or $\frac{j+1}{2k-1}$, each with probability $0.5$. 
	Denote $x_i = \frac{j}{2k-1} + \epsilon$ for some $0 \leq \epsilon < \frac{1}{2k-1}$. 
	The expected cost of the agent will be:
	$0.5\left(|x_i-\frac{j}{2k-1}|+|\frac{j+1}{2k-1}-x_i|\right) = 0.5(\epsilon + \frac{1}{2k-1} - \epsilon) = \frac{1}{4k-2}$.
\end{proof}

For the deterministic case we introduce election-parity-equal-cost (EPEC). Like in PEC, facilities there are two options - locating facilities on $\{\frac{i}{2k-1}\}$ for even or odd values of $i$. 
In order to decide whether to choose the even or odd values, the mechanism counts the amount of agents who prefer each option (based on their reports), and decides based on majority (ties can be broken arbitrarily).

\begin{theorem} \label{thm:epec}
	EPEC is truthful, and achieves an error of $\frac{3}{8k-4}$ for the average cost.
\end{theorem}

\begin{proof} 
	EPEC is truthful --- the agents have 2 options to choose from, and the decision is done based on majority, therefore no agent has an incentive to misreport to a position which prefers the other option. 
	Any misreport which remains within the same option (for instance, declaring $x_i'$ at location $x_i$, where both $x_i,x_i'$ are closest to $\frac{i}{2k-1}$ for even $i$) does not affect the result.
	
	The cost of any agent who voted for the option which was chosen is at most $\frac{1}{2(2k-1)}$. 
	The cost of any agent is at most $\frac{1}{2k-1}$.
	Since we chose based on majority, the error is at most: $$\frac{1}{n}\left[  \frac{n}{2} \left( \frac{1}{4k-2} \right) +  \frac{n}{2} \left(\frac{1}{2k-1} \right) \right] = \frac{3}{8k-4}$$
\end{proof}

For the case of two facilities ($k=2$), we can further improve the result of EPEC:
\begin{definition} [Fifths mechanism]
	Let $\bx$ be the location profile in ascending order. The fifths mechanism locates the 2 facilities on reports $x_{n/5}$ and $x_{4n/5}$.
\end{definition}

\begin{theorem} \label{thm:fifths}
	The fifths mechanism is truthful and has an error $\frac{1}{5}$ for the average cost.
\end{theorem}

The proof is deferred to the appendix.

\section{Discussion and Open Problems}

We examine the problem of truthful facility location under additive errors. We proved tight bounds for one facility, and showed several extensions for multiple facilities. 

We believe that there are many interesting directions which were explored for the multiplicative error but remain wide open for the additive error. For instance, exploring additional metric spaces (cycles, trees, etc.), different objective functions (e.g., $L_2$) or other cost functions of the agents (e.g. ``obnoxious facility location". In addition, other models can also be researched with this error function in mind, for example voting via mediators (see \cite{babaioff2016mechanism}) or when voting for a discrete set of candidates (see \cite{feldman2015voting}). 

\paragraph{Acknowledgements.}
We wish to thank Prof. Michal Feldman and Prof. Amos Fiat for their helpful suggestions.

\bibliographystyle{plain}
\bibliography{refs}

\begin{thebibliography}{10}

\bibitem{alon2009strategyproof}
Noga Alon, Michal Feldman, Ariel~D Procaccia, and Moshe Tennenholtz.
\newblock Strategyproof approximation mechanisms for location on networks.
\newblock {\em arXiv preprint arXiv:0907.2049}, 2009.

\bibitem{alon2010walking}
Noga Alon, Michal Feldman, Ariel~D Procaccia, and Moshe Tennenholtz.
\newblock Walking in circles.
\newblock {\em Discrete Mathematics}, 310(23):3432--3435, 2010.

\bibitem{anshelevich2015approximating}
Elliot Anshelevich, Onkar Bhardwaj, and John Postl.
\newblock Approximating optimal social choice under metric preferences.
\newblock In {\em AAAI}, volume~15, pages 777--783. Citeseer, 2015.

\bibitem{anshelevich2015randomized}
Elliot Anshelevich and John Postl.
\newblock Randomized social choice functions under metric preferences.
\newblock {\em arXiv preprint arXiv:1512.07590}, 2015.

\bibitem{babaioff2016mechanism}
Moshe Babaioff, Moran Feldman, and Moshe Tennenholtz.
\newblock Mechanism design with strategic mediators.
\newblock {\em ACM Transactions on Economics and Computation}, 4(2):7, 2016.

\bibitem{caifacility}
Qingpeng Cai, Aris Filos-Ratsikas, Aris Filos, and Pingzhong Tang.
\newblock Facility location with minimax envy.

\bibitem{cheng2013strategy}
Yukun Cheng, Wei Yu, and Guochuan Zhang.
\newblock Strategy-proof approximation mechanisms for an obnoxious facility
  game on networks.
\newblock {\em Theoretical Computer Science}, 497:154--163, 2013.

\bibitem{dokow2012mechanism}
Elad Dokow, Michal Feldman, Reshef Meir, and Ilan Nehama.
\newblock Mechanism design on discrete lines and cycles.
\newblock In {\em Proceedings of the 13th ACM Conference on Electronic
  Commerce}, pages 423--440. ACM, 2012.

\bibitem{feigenbaum2015strategyproof}
Itai Feigenbaum and Jay Sethuraman.
\newblock Strategyproof mechanisms for one-dimensional hybrid and obnoxious
  facility location models.
\newblock In {\em Workshops at the Twenty-Ninth AAAI Conference on Artificial
  Intelligence}, 2015.

\bibitem{feldman2015voting}
Michal Feldman, Amos Fiat, and Iddan Golomb.
\newblock On voting and facility location.
\newblock In {\em Proceedings of the 2016 {ACM} Conference on Economics and
  Computation, {EC} '16, Maastricht, The Netherlands, July 24-28, 2016}, pages
  269--286, 2016.

\bibitem{feldman2013strategyproof}
Michal Feldman and Yoav Wilf.
\newblock Strategyproof facility location and the least squares objective.
\newblock In {\em Proceedings of the fourteenth ACM conference on Electronic
  commerce}, pages 873--890. ACM, 2013.

\bibitem{filos2015facility}
Aris Filos-Ratsikas, Minming Li, Jie Zhang, and Qiang Zhang.
\newblock Facility location with double-peaked preferences.
\newblock In {\em Proceedings of the Twenty-Ninth AAAI Conference on Artificial
  Intelligence}, pages 893--899. AAAI Press, 2015.

\bibitem{fotakis2010winner}
Dimitris Fotakis and Christos Tzamos.
\newblock Winner-imposing strategyproof mechanisms for multiple facility
  location games.
\newblock In {\em International Workshop on Internet and Network Economics},
  pages 234--245. Springer, 2010.

\bibitem{fotakis2013strategyproof}
Dimitris Fotakis and Christos Tzamos.
\newblock Strategyproof facility location for concave cost functions.
\newblock In {\em Proceedings of the fourteenth ACM conference on Electronic
  commerce}, pages 435--452. ACM, 2013.

\bibitem{fotakis2014power}
Dimitris Fotakis and Christos Tzamos.
\newblock On the power of deterministic mechanisms for facility location games.
\newblock {\em ACM Transactions on Economics and Computation}, 2(4):15, 2014.

\bibitem{gross2017vote}
Stephen Gross, Elliot Anshelevich, and Lirong Xia.
\newblock Vote until two of you agree: Mechanisms with small distortion and
  sample complexity.
\newblock 2017.

\bibitem{groves1973incentives}
Theodore Groves.
\newblock Incentives in teams.
\newblock {\em Econometrica: Journal of the Econometric Society}, pages
  617--631, 1973.

\bibitem{lu2010asymptotically}
Pinyan Lu, Xiaorui Sun, Yajun Wang, and Zeyuan~Allen Zhu.
\newblock Asymptotically optimal strategy-proof mechanisms for two-facility
  games.
\newblock In {\em Proceedings of the 11th ACM conference on Electronic
  commerce}, pages 315--324. ACM, 2010.

\bibitem{lu2009tighter}
Pinyan Lu, Yajun Wang, and Yuan Zhou.
\newblock Tighter bounds for facility games.
\newblock In {\em Internet and Network Economics}, pages 137--148. Springer,
  2009.

\bibitem{moulin1980strategy}
Herv{\'e} Moulin.
\newblock On strategy-proofness and single peakedness.
\newblock {\em Public Choice}, 35(4):437--455, 1980.

\bibitem{nissim2012approximately}
Kobbi Nissim, Rann Smorodinsky, and Moshe Tennenholtz.
\newblock Approximately optimal mechanism design via differential privacy.
\newblock In {\em Proceedings of the 3rd innovations in theoretical computer
  science conference}, pages 203--213. ACM, 2012.

\bibitem{procaccia2006distortion}
Ariel~D Procaccia and Jeffrey~S Rosenschein.
\newblock The distortion of cardinal preferences in voting.
\newblock In {\em International Workshop on Cooperative Information Agents},
  pages 317--331. Springer, 2006.

\bibitem{procaccia2009approximate}
Ariel~D Procaccia and Moshe Tennenholtz.
\newblock Approximate mechanism design without money.
\newblock In {\em Proceedings of the 10th ACM conference on Electronic
  commerce}, pages 177--186. ACM, 2009.

\bibitem{procacciaapproximation}
Ariel~D Procaccia, David Wajc, and Hanrui Zhang.
\newblock Approximation-variance tradeoffs in mechanism design.

\bibitem{schummer2002strategy}
James Schummer and Rakesh~V Vohra.
\newblock Strategy-proof location on a network.
\newblock {\em Journal of Economic Theory}, 104(2):405--428, 2002.

\bibitem{serafino2015truthful}
Paolo Serafino and Carmine Ventre.
\newblock Truthful mechanisms without money for non-utilitarian heterogeneous
  facility location.
\newblock In {\em AAAI}, pages 1029--1035, 2015.

\bibitem{sui2015approximately}
Xin Sui and Craig Boutilier.
\newblock Approximately strategy-proof mechanisms for (constrained) facility
  location.
\newblock In {\em Proceedings of the 2015 International Conference on
  Autonomous Agents and Multiagent Systems}, pages 605--613. International
  Foundation for Autonomous Agents and Multiagent Systems, 2015.

\bibitem{zou2015facility}
Shaokun Zou and Minming Li.
\newblock Facility location games with dual preference.
\newblock 2015.

\end{thebibliography}

\includepdf[pages=-]{appendix}

\end{document}


\setcounter{page}{8}
\pagenumbering{gobble}
\appendix
\section{Analysis of the Fifths mechanism}
Both facilities are chosen based on the percentiles of the agents. The mechanism is truthful -- for each agent and each of the two facilities, misreporting can either not affect the location of the facility, or cause it to be located farther from the agent.
	
	For a given location profile, if the amount of agents is not a multiple of 5, we analyze the case in which we duplicate the location of each agent 5 times. As shown in Observation 1,
  this does not affect the additive error. We denote the new location profile by $\bx$.
	
	
	Let $OPT_1 < OPT_2$ be the locations of the two facilities in some optimal mechanism. 
	Let $F_1=x_{1/5}, F_2=x_{4n/5}$ be the locations of the two facilities in the fifths mechanism.
	
	We split the proof to different cases:
	\begin{enumerate}
		\item $OPT_1 \in [0,F_1)$ and $OPT_2 \in (F_2,1]$: We show that this case is not possible. 
		
		In the single facility case, the optimal mechanism locates the facility on the median. Therefore, in the two facility case the optimal mechanism divide the agents into two groups: $A=\{x_1,\ldots x_k\}$ and $B=\{x_{k+1}\ldots x_n \}$, and locates the $OPT_1,OPT_2$ on the median of groups $A$ and $B$ respectively. 
		
		Therefore, either $A$ or $B$ contains at least half of the agents. Assume without loss of generality that $|A| \geq n/2$. Therefore there exist at least $n/4$ agents to the left of the median of $OPT_1$, so $OPT_1 \geq F_1 \Rightarrow OPT_1 \notin[0,F_1)$.
				
		\item If $OPT_1,OPT_2 \in [F_1,F_2]$: In this case we show a series of transitions from $\bx$ to a different location profile (without changing the location of $F_1$ and $F_2$) whose additive error can only be greater. We show that the additive error of the new profile is not greater than $0.2$.

		We now show the transitions, and prove that in each case the difference in the additive error is non-negative. 
		In each step $i$ we denote the current location profile by $\bx^{i}$ (in the initial stage:$\bx^{0}=\bx$). 
		For some agent $j$ let $\beta^i_j,\gamma^i_j$ be the difference in the costs of the agent at stage $i$ under $M$ and under $OPT$ respectively, that is: $\beta^i_j=\cost_{x_j}(M,\bx^i)-\cost_{x_j}(M,\bx^{i-1})$ and 
		$\gamma^i_j=\cost_{x_j}(OPT,\bx^i)-\cost_{x_j}(OPT,\bx^{i-1})$.
		The total difference in the additive error in step $i$ is denoted by $\Delta^{i}=\sum_j(\beta^i_j-\gamma^i_j)$.
		We show that $\forall i: \Delta^{i} \geq 0$. We measure the error based on the fixed locations $OPT_1,OPT_2$, and if the optimal mechanism changes, this makes  difference in additive error even higher.
		
		Let $C = \frac{F_1+F_2}{2}$, and assume without loss of generality that $|C-OPT_1|\geq |C-OPT_2|$. 
		We analyze the following transitions (see Figure \ref{fig:fifths-first-case-process}):
		\begin{enumerate}
			\item $\bx^{1}$: Move all agents in $[0,F_1)$ to $F_1$: In this case all agents which move get closer to $OPT_1$ by the same amount that they get closer to $F_1$, that is $\beta^1_j-\gamma^1_j$.  
			Let:
			\begin{align*}				
				x^1_j = \begin{cases}
					F_1 & \quad \text{if }  x_j \in [0,F_1) \\
					x_j & \quad \text{otherwise } 
				\end{cases}
			\end{align*}
			The change in the additive error is:
			$\Delta^{1}=\sum_{j}(\beta^1_j-\gamma^1_j)=\sum_{j:x_j<F_1}(\beta^1_j-\gamma^1_j)=0$.
			
			\item $\bx^2$: Move all agents in $(F_2,1]$ to $F_2$: In this case all agents which move get closer to $OPT_2$ by the same amount that they get closer to $F_2$, that is $\beta^2_j-\gamma^2_j$. 
			Let:
			\begin{align*}				
			x^2_j = \begin{cases}
			F_2 & \quad \text{if }  x^1_j \in (F_2,1] \\
			x_j^1 & \quad \text{otherwise } 
			\end{cases}
			\end{align*}
			The change in the additive error is:
			$\Delta^{2}=\sum_{j}(\beta^2_j-\gamma^2_j)=\sum_{j:x_j>F_2}(\beta^2_j-\gamma^2_j)=0$.

			\item $\bx^3$: Move all agents in $(F_1,F_2)$ to $OPT_2$: In this case each agent $j$ which moves distance $\delta_j$, reduces the cost of $OPT$ by $\delta_j$ and reduces the cost of $M$ by no more than $\delta_j$, therefore $\forall j: \beta^3_j \geq \gamma^3_j$.
			Let:
			\begin{align*}				
			x^3_j = \begin{cases}
			OPT_2 & \quad \text{if }  x^2_j \in (F_1,F_2) \\
			x_j^2 & \quad \text{otherwise } 
			\end{cases}
			\end{align*}
			The change in the additive error is:
			$\Delta^{3}=\sum_{j}(\beta^3_j-\gamma^3_j)=\sum_{j:F_1<x_j<F_2}(\beta^3_j-\gamma^3_j)\geq 0$.	
						
			\item $\bx^4$: Scale the segment $[F_1,F_2]$ to $[0,1]$. Notice that in $\bx^3$ all agents are within the segment $[F_1,F_2]$. The scaling increases all the distances between agents by the same factor, so the additive error of any agent will also increase (by that factor): $\forall j: \beta^4_j - \gamma^4_j\geq 0$. Let $D=\frac{1}{F_2-F_1} \geq 1$ and let:
			\begin{align*}				
			x^4_j =  \frac{x^3_j-F_1}{F_2-F_1}
			\end{align*}
			For every agent $j$ it holds that $\beta^4_j=D\cdot \beta^3_j$ and $\gamma^4_j=D\cdot \gamma^3_j$. The change in the additive error is therefore:
			$\Delta^{4}=\sum_{j}(\beta^4_j-\gamma^4_j)=D \cdot \sum_{j}(\beta^3_j-\gamma^3_j)\geq 0$.			
		\end{enumerate}

		\begin{figure}[h]		
			\begin{framed}
				\caption{The transitions from profile $\bx$ to $\bx^3$:}							
				
				{\begin{minipage}{1\textwidth}				
						\centering							
						\begin{tikzpicture}[y=.3cm, x=.3cm,font=\sffamily]			
						\draw[<->, thick] (0,0) -- (40,0) node{};
						
						\draw[fill=black] (0, 0) circle (0.4) node [below, yshift=-0.3cm] {};
						\draw[fill=black] (7, 0) circle (0.4) node [below, yshift=-0.3cm] {$F_1$};		
						\draw[fill=black] (11, 0) circle (0.4) node [below, yshift=-0.3cm] {$OPT_1$};
						\draw[fill=black] (14, 0) circle (0.4) node [below, yshift=-0.3cm] {};
						\draw[fill=black] (17, 0) circle (0.4) node [below, yshift=-0.3cm] {};
						\draw[fill=black] (20, 0) circle (0.4) node [below, yshift=-0.3cm] {$OPT_2$};	
						\draw[fill=black] (23, 0) circle (0.4) node [below, yshift=-0.3cm] {};			
						\draw[fill=black] (28, 0) circle (0.4) node [below, yshift=-0.3cm] {$F_2$};
						\draw[fill=black] (32, 0) circle (0.4) node [below, yshift=-0.3cm] {};			

						\draw[->, thick] (0.4,0.5) .. controls (3.5,1.5) ..(6.7,0.5);	
						
						\draw[->, thick] (11.4,0.5) .. controls (15,1.5) ..(19.6,0.5);							
						\draw[->, thick] (14.4,0.5) .. controls (17,0.8) ..(19.6,0.5);	
						\draw[->, thick] (17.4,0.5) .. controls (18.5,0.6) ..(19.6,0.5);																			
						
						\draw[->, thick] (22.6,0.5) .. controls (21.5,0.8) ..(20.4,0.5);													

						\draw[->, thick] (31.6,0.5) .. controls (30,0.8) ..(28.4,0.5);													
						
						
						%
						
						\end{tikzpicture}

					\end{minipage}} 
					\label{fig:fifths-first-case-process}
				\end{framed}
			\end{figure}

		At the end of the process the profile is $\bx^4$. The highest additive error in $\bx^4$ is reached in the following configuration (see Figure \ref{fig:fifths-first-case}): when there are $n/5$ agents at $0$, $3n/5$ agents at some point $x$ (the location of $OPT_2$), and $n/5$ agents at point $1$. We now assess the error with respect to the new assignment of $OPT$ (that is, moving $OPT_1$ to $0$), which can only increase the additive error.
		Assume without loss of generality that $x \geq 1/2$.
		
		In this case, $OPT_1=0,OPT_2=x$. The costs are $AC(OPT)=\frac{1-x}{5}$ and $AC(M)=\frac{3(1-x)}{5}$.
		Since $x\geq 1/2$ the additive error is: $\frac{3(1-x)}{5}-\frac{1-x}{5}=\frac{2(1-x)}{5}\leq \frac{1}{5}$.
		
		\begin{figure}[h]		
			\begin{framed}
				\caption{The location profile $\bx^4$}							
				
				{\begin{minipage}{1\textwidth}				
					\centering							
					\begin{tikzpicture}[y=.3cm, x=.3cm,font=\sffamily]			
					\draw[<->, thick] (0,0) -- (40,0) node{};

					\draw[fill=black] (0, 0) circle (0.4) node [below, yshift=0.7cm] {$0$};
					\draw[fill=black] (0, 0) circle (0.4) node [below, yshift=-0.3cm] {$F_1=OPT_1$};
					\draw[fill=black] (0, 0) circle (0.4) node [below, yshift=-0.7cm] {$n/5$};								
					
					\draw[fill=black] (23, 0) circle (0.4) node [below, yshift=0.7cm] {$x$};
					\draw[fill=black] (23, 0) circle (0.4) node [below, yshift=-0.3cm] {$OPT_2$};
					\draw[fill=black] (23, 0) circle (0.4) node [below, yshift=-0.7cm] {$3n/5$};									
					
					\draw[fill=black] (40, 0) circle (0.4) node [below, yshift=0.7cm] {$1$};										
					\draw[fill=black] (40, 0) circle (0.4) node [below, yshift=-0.3cm] {$F_2$};			
					\draw[fill=black] (40, 0) circle (0.4) node [below, yshift=-0.7cm] {$n/5$};									
					
%
					
					\end{tikzpicture}

				\end{minipage}} 
				\label{fig:fifths-first-case}
			\end{framed}
		\end{figure}

		\item If $OPT_1 \in [0,F_1)$ and $OPT_2 \in [F_1,F_2]$.
		
		Similarly to the previous case, we show a series of transitions which does not decrease the additive error (see Figure \ref{fig:fifths-second-case-process}). 
		We denote the location profiles with brackets (e.g., $\bx^{(1)}$) do distinguish the notation from the previous case.
		\begin{enumerate}
			\item $\bx^{(1)}$: Move all agents in $[0,F_1)$ to $OPT_1$: In this case the additive error does not change for all agents in $[0,OPT_1] \cup [F_1,1]$, and the error increases for agents in $(OPT_1,F_1)$.
			Let:
			\begin{align*}				
				x^{(1)}_j = \begin{cases}
					OPT_1 & \quad \text{if }  x_j \in [0,F_1) \\
					x_j & \quad \text{otherwise } 
				\end{cases}
			\end{align*}
			The change in the additive error is:
			$\Delta^{(1)}=\sum_{j}(\beta^{(1)}_j-\gamma^{(1)}_j)=\sum_{j:OPT_1<x_j<F_1}(\beta^{(1)}_j-\gamma^{(1)}_j)\geq 0$.

			\item $\bx^{(2)}$: Move all agents in $(F_1,F_2)$ to $OPT_2$: In this case each agent $j$ which moves distance $\delta_j$ to $OPT_2$, reduces the cost of $OPT$ by $\delta_j$ and reduces the cost of $M$ by no more than $\delta_j$, therefore $\forall j: \beta^{(2)}_j \geq \gamma^{(2)}_j$.
			Let:
			\begin{align*}				
				x^{(2)}_j = \begin{cases}
					OPT_2 & \quad \text{if }  x^{(1)}_j \in (F_1,F_2) \\
					x^{(1)}_j & \quad \text{otherwise } 
				\end{cases}
			\end{align*}
			The change in the additive error is:
			$\Delta^{(2)}=\sum_{j}(\beta^{(2)}_j-\gamma^{(2)}_j)=\sum_{j:F_1<x_j<F_2}(\beta^{(2)}_j-\gamma^{(2)}_j)\geq 0$.
			
			\item $\bx^{(3)}$: Move all agents in $(F_2,1]$ to $F_2$. In this case all agents which move get closer to $OPT_2$ by the same amount that they get closer to $F_2$, that is $\beta^2_j-\gamma^2_j$. 
			Let:
			\begin{align*}				
				x^{(3)}_j = \begin{cases}
					F_2 & \quad \text{if }  x^{(2)}_j \in (F_2,1] \\
					x^{(2)}_j & \quad \text{otherwise } 
				\end{cases}
			\end{align*}
			The change in the additive error is:
			$\Delta^{(3)}=\sum_{j}(\beta^{(3)}_j-\gamma^{(3)}_j)=\sum_{j:F_2<x_j\leq 1}(\beta^{(3)}_j-\gamma^{(3)}_j)= 0$.

\begin{figure}[t]		
	\begin{framed}
		\caption{The transitions from profile $\bx$ to $\bx^{(3)}$:}							
		
		{\begin{minipage}{1\textwidth}				
				\centering							
				\begin{tikzpicture}[y=.3cm, x=.3cm,font=\sffamily]			
				\draw[<->, thick] (0,0) -- (40,0) node{};
				
				\draw[fill=black] (0, 0) circle (0.4) node [below, yshift=-0.3cm] {};
				\draw[fill=black] (7, 0) circle (0.4) node [below, yshift=-0.3cm] {$OPT_1$};
				\draw[fill=black] (10, 0) circle (0.4) node [below, yshift=-0.3cm] {};								
				\draw[fill=black] (12, 0) circle (0.4) node [below, yshift=-0.3cm] {$F_1$};
				\draw[fill=black] (14, 0) circle (0.4) node [below, yshift=-0.3cm] {};
				\draw[fill=black] (17, 0) circle (0.4) node [below, yshift=-0.3cm] {};
				\draw[fill=black] (20, 0) circle (0.4) node [below, yshift=-0.3cm] {$OPT_2$};	
				\draw[fill=black] (23, 0) circle (0.4) node [below, yshift=-0.3cm] {};			
				\draw[fill=black] (28, 0) circle (0.4) node [below, yshift=-0.3cm] {$F_2$};
				\draw[fill=black] (32, 0) circle (0.4) node [below, yshift=-0.3cm] {};			
				
				\draw[->, thick] (0.4,0.5) .. controls (3.5,1.5) ..(6.7,0.5);	
				
				\draw[->, thick] (9.6,0.5) .. controls (8.5,0.7) ..(7.4,0.5);	
				
				\draw[->, thick] (14.4,0.5) .. controls (17,0.8) ..(19.6,0.5);	
				\draw[->, thick] (17.4,0.5) .. controls (18.5,0.6) ..(19.6,0.5);																			
				
				\draw[->, thick] (22.6,0.5) .. controls (21.5,0.8) ..(20.4,0.5);													
				
				\draw[->, thick] (31.6,0.5) .. controls (30,0.8) ..(28.4,0.5);													
				
				
				%
				
				\end{tikzpicture}

			\end{minipage}} 
			\label{fig:fifths-second-case-process}
		\end{framed}
	\end{figure}
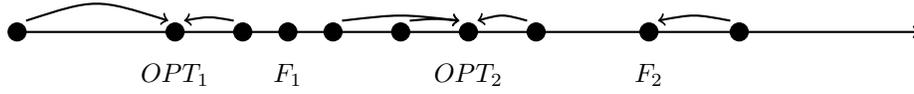

			\item $\bx^{(4)}$: Scale $[OPT_1,F_2]$ to $[0,1]$. Notice that in $\bx^{(3)}$ all agents are within the segment $[OPT_1,F_2]$. 
			The scaling increases all the distances between agents by the same factor, so the additive error of any agent will also increase (by that factor): $\forall j: \beta^{(4)}_j - \gamma^{(4)}_j\geq 0$. Let $D'=\frac{1}{F_2-OPT_1} \geq 1$ and let:
			\begin{align*}				
				x^{(4)}_j =  \frac{x^{(3)}_j-OPT_1}{F_2-OPT_1}
			\end{align*}
			For every agent $j$ it holds that $\beta^{(4)}_j=D'\cdot \beta^{(3)}_j$ and $\gamma^{(4)}_j=D'\cdot \gamma^{(3)}_j$. The change in the additive error is therefore:
			$\Delta^{(4)}=\sum_{j}(\beta^{(4)}_j-\gamma^{(4)}_j)=D' \cdot \sum_{j}(\beta^{(4)}_j-\gamma^{(4)}_j)\geq 0$.				
			
		\end{enumerate}

		\begin{figure}[h]		
			\begin{framed}
				\caption{The location profile $\bx^{(4)}$}							
				
				{\begin{minipage}{1\textwidth}				
						\centering							
						\begin{tikzpicture}[y=.3cm, x=.3cm,font=\sffamily]			
						\draw[<->, thick] (0,0) -- (40,0) node{};
						
						\draw[fill=black] (0, 0) circle (0.4) node [below, yshift=0.7cm] {$0$};
						\draw[fill=black] (0, 0) circle (0.4) node [below, yshift=-0.3cm] {$OPT_1$};
						\draw[fill=black] (0, 0) circle (0.4) node [below, yshift=-0.7cm] {$n/5$};	
						
						\draw[fill=black] (7, 0) circle (0.4) node [below, yshift=-0.3cm] {$F_1$};
						\draw[fill=black] (7, 0) circle (0.4) node [below, yshift=-0.7cm] {$1$};							
						\draw[dim] (0,3) -- (7,3) node[midway,above] {$1-x-y$};

						\draw[fill=black] (25, 0) circle (0.4) node [below, yshift=-0.3cm] {$OPT_2$};
						\draw[fill=black] (25, 0) circle (0.4) node [below, yshift=-0.7cm] {$3n/5$};									
						\draw[dim] (7,3) -- (25,3) node[midway,above] {$y$};
						
						\draw[fill=black] (40, 0) circle (0.4) node [below, yshift=0.7cm] {$1$};										
						\draw[fill=black] (40, 0) circle (0.4) node [below, yshift=-0.3cm] {$F_2$};			
						\draw[fill=black] (40, 0) circle (0.4) node [below, yshift=-0.7cm] {$n/5$};	
						\draw[dim] (25,3) -- (40,3) node[midway,above] {$x$};

						%
						
						\end{tikzpicture}

					\end{minipage}} 
					\label{fig:fifths-second-case}
				\end{framed}
			\end{figure}

		At the end of the process, the highest additive error is reached in $\bx^{(4)}$ in the following configuration (see Figure \ref{fig:fifths-second-case}) when there are $n/5$ agents at point $0$ (location of $OPT_1$), one agent at $F_1$, $3n/5$ agents at $OPT_2$ and $n/5$ agents at point $1$ (location of $F_2$). We denote the distances $x=F_2-OPT_2$, $y=OPT_2-F_1$ and therefore $F_1=1-x-y$. Denote the cost of the agent at $F_1$ by $\alpha$.
		
		In this case $\cost(OPT)=x/5 + \alpha \geq x/5$ and $\cost(M)= \frac{3\cdot \min\{x,y\}}{5} + \frac{1-x-y}{5}$. Therefore the additive error is $\frac{1}{5}\left( 1+3\min\{x,y\} - (2x+y) \right) \leq 1/5$.

		\item If $OPT_1 \in [F_1,F_2]$ and $OPT_2 \in (F_2,1]$: Completely symmetric to the previous case.
	\end{enumerate}